\documentclass[runningheads]{llncs}
\usepackage[T1]{fontenc}
\usepackage{microtype}
\usepackage{graphicx}
\usepackage{subfigure}
\usepackage{booktabs} 

\usepackage{hyperref}

\usepackage{amsmath}
\usepackage{amssymb}
\usepackage{mathtools}

\usepackage[capitalize,noabbrev]{cleveref}

\usepackage[textsize=tiny]{todonotes}

\usepackage{verbatim,float,url,enumerate}
\usepackage{graphicx,subfigure,epsfig,psfrag}
\usepackage{dsfont,bm,color,appendix}
\usepackage{natbib}
\usepackage{pdfpages}
\usepackage{algorithm}
\usepackage{algpseudocode}
\usepackage{bbm}
\usepackage[algo2e]{algorithm2e}
\usepackage{enumitem}
\usepackage[thinc]{esdiff}
\usepackage{afterpage}
\usepackage{amsmath,amssymb,bm}
\usepackage{booktabs}
\usepackage{multirow}
\usepackage{array}
\usepackage{xcolor}
\usepackage{physics}
\usepackage[symbol]{footmisc}
\usepackage[font=footnotesize,labelfont=bf]{caption}

\newcommand\blfootnote[1]{%
  \begingroup
  \renewcommand\thefootnote{}\footnote{#1}%
  \addtocounter{footnote}{-1}%
  \endgroup
}

\newcommand{\mR}{\mathcal{R}}

\newcommand{\be}{\begin{equation}}
\newcommand{\ee}{\end{equation}}
\renewcommand{\vec}[1]{\bm{#1}}
\newcommand{\tx}{\vec{\theta_x}}

\newcommand{\tz}{\vec{\theta_z}}

\newcommand{\bbeta}{\vec{\beta}}
\newcommand{\y}{\vec{y}}

\newcolumntype{M}[1]{>{\centering\arraybackslash}m{#1}}
\renewcommand{\thefootnote}{\fnsymbol{footnote}}

\usepackage{graphicx}
\begin{document}
\title{Semi-supervised Cooperative Learning for Multiomics Data Fusion}
%
%
\author{Daisy Yi Ding \and
Xiaotao Shen \and
Michael Snyder \and
Robert Tibshirani}
\authorrunning{Ding et al.}
%
\institute{Stanford University, Stanford, CA 94305, USA} 
%
\maketitle              
\begin{abstract}
Multiomics data fusion integrates diverse data modalities, ranging from transcriptomics to proteomics, to gain a comprehensive understanding of biological systems and enhance predictions on outcomes of interest related to disease phenotypes and treatment responses.
Cooperative learning, a recently proposed method, unifies the commonly-used fusion approaches, including early and late fusion, and offers a systematic framework for leveraging the shared underlying relationships across omics to strengthen signals.
However, the challenge of acquiring large-scale labeled data remains, and there are cases where multiomics data are available but in the absence of annotated labels.
To harness the potential of unlabeled multiomcis data, we introduce semi-supervised cooperative learning.
By utilizing an ``agreement penalty'', our method incorporates the additional unlabeled data in the learning process
and achieves consistently superior predictive performance on simulated data and a real multiomics study of aging.
It offers an effective solution to multiomics data fusion in settings with both labeled and unlabeled data and maximizes the utility of available data resources, with the potential of significantly improving predictive models for diagnostics and therapeutics in an increasingly multiomics world. \blfootnote{\textit{The 2023 ICML Workshop on Machine Learning for Multimodal Healthcare Data.}}

\keywords{Multiomics data fusion  \and Semi-supervised learning \and Machine learning.}
\end{abstract}

\section{Introduction}

With advancements in biotechnologies, significant progress has been made in generating and collecting a diverse range of ``-omics'' data on a common set of patients, including genomics, epigenomics, transcriptomics, proteomics, and metabolomics (Figure 1{\em A}). 
These data characterize molecular variations of human health from different perspectives and of different granularities.
Fusing the multiple data modalities on a common set of observations provides the opportunity to gain a more holistic understanding of outcomes of interest such as disease phenotypes and treatment response. 
It offers the potential to discover hidden insights that may remain obscured in single-modality data analyses and achieve more accurate predictions of the outcomes \citep{kristensen2014principles, ritchie2015methods, robinson2017integrative, karczewski2018integrative, ma2020integrative, hao2021integrated}. 
While the term ``multiomics data fusion'' can have various interpretations, we use it here in the context of predicting an outcome of interest by integrating different data modalities.

\begin{figure*}[t!]
    \centering
    \includegraphics[width=1\textwidth]{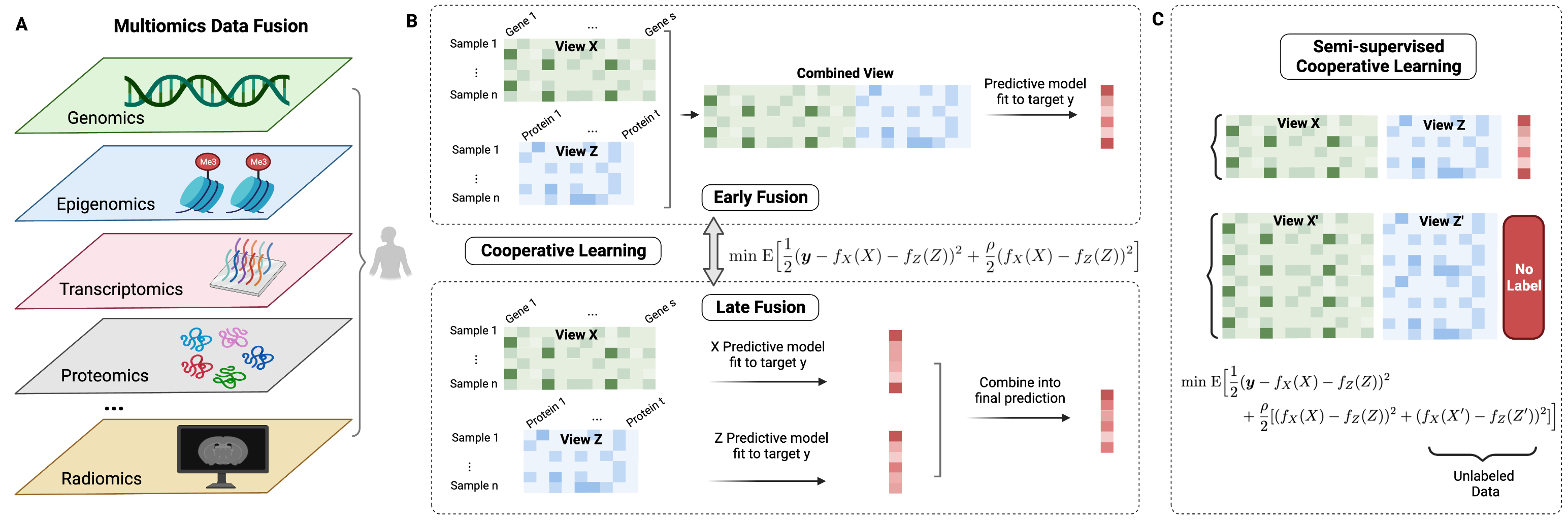}
    \caption{\emph{Framework of semi-supervised cooperative learning for multiomics data fusion.} \emph{(A)} The advancements in biotechnologies have led to the generation and collection of diverse ``omics'' data on a common set of samples, ranging from genomics to proteomics.
    Fusing the data provides a unique opportunity to gain a holistic understanding of complex biological systems and enhance predictive accuracy on outcomes of interest related to disease phenotypes and treatment response
\emph{(B)} Commonly-used approaches to multiomics data fusion have two broad categories: early fusion involves transforming all data modalities into a unified representation before feeding it into a model of choice, while late fusion builds separate models for each data modality and combines their predictions using a second-level model.
Encompassing early and late fusion, 
cooperative learning exploits the shared underlying relationships across omics for enhanced predictive performance. 
\emph{(C)} The field of biomedicine faces a persistent challenge due to the scarcity of large-scale labeled data, which requires significant resources to acquire. 
In cases where unlabeled multiomics data are also accessible, we introduce semi-supervised cooperative learning to leverage the combined information from both labeled and unlabeled data.
The agreement penalty seamlessly integrates the unlabeled samples into the learning process, effectively utilizing the shared underlying signals present in both labeled and unlabeled data and maximizing the utility of available data resources for multiomics data fusion.
}
\end{figure*}

Commonly-used data fusion methods can be broadly categorized into early and late fusion (Figure 1{\em B}).
Early fusion works by transforming the multiple data modalities into a single representation before feeding the aggregated representation into a supervised learning model of choice \citep{yuan2014assessing,gentles2015integrating,perkins2018precision,chaudhary2018deep}.
Late fusion refers to methods where individual models are first built from the distinct data modality, and then the predictions of the individual models are combined into the final predictor \citep{yang2010review, zhao2019learning, chen2020pathomic, chabon2020integrating, wu2020integrative}. 
However, both methods do not explicitly leverage the shared underlying relationship across modalities, and a systematic framework for multiomics data fusion is lacking.

To tackle this limitation, a new method called \emph{cooperative learning} has recently been proposed \citep{ding2022cooperative}.
It combines the usual squared error loss of predictions with an  ``agreement'' penalty to encourage alignment of predictions from different data modalities (Figure 1{\em B}). 
By varying the weight of the agreement penalty, one can get a continuum of solutions that include early and late fusion. 
Cooperative learning chooses the degree of fusion in a data-adaptive manner, providing enhanced flexibility and performance.
It has demonstrated effectiveness on both simulated data and real multiomics data, particularly when 
the different data modalities share some underlying relationships in their signals that can be exploited to boost the signals. 

However, an important challenge persists in the field of biomedicine: the scarcity of large-scale labeled data.
Acquiring a substantial amount of labeled data in this domain often demands considerable effort, time, and financial resources.
Nonetheless, there are instances where multiomics data are available, but in the absence of corresponding labels. 
In such cases, it becomes imperative to leverage the available unlabeled data to enhance predictive models.

To harness the potential of unlabeled data, we propose \emph{semi-supervised cooperative learning}. 
The key idea is to utilize the agreement penalty, inherent in the cooperative learning framework, as a means to leverage the matched unlabeled samples to our advantage (Figure 1{\em C}). 
It acts as a mechanism for incorporating the unlabeled samples into the learning process, by encouraging the predictions from different data modalities to align not only on the labeled samples but also on the unlabeled ones. 
Semi-supervised cooperative learning leverages the additional shared underlying signals across the unlabeled data and exploits the valuable information that would otherwise remain untapped.
Through comprehensive simulated studies and a real multiomics study of aging, we showed that our method achieves consistently higher predictive accuracy on the outcomes of interest.
By incorporating matched unlabeled data and thus maximizing the utility of available data, semi-supervised cooperative learning offers an effective solution to multiomics data fusion, with the potential to significantly enhance predictive models and unlock hidden insights in health and disease.


\section{Approach}

\subsection{Cooperative learning}

We begin by giving a concise overview of the recently proposed \emph{cooperative learning} framework \citep{ding2022cooperative} to set the stage for the introduction of \emph{semi-supervised cooperative learning}.
Let $X \in
\mR^{n \times p_x}$, $Z \in
\mR^{n \times p_z}$ --- representing two data views --- and  $\vec{y} \in \mR^{n}$ be a real-valued response. Fixing the
hyperparameter $\rho\geq 0$, cooperative learning aims to minimize the population quantity:
\begin{equation}
{\rm min} \;  {\rm E}\Bigl[\frac{1}{2}  (\y-f_X(X)-f_Z(Z))^2+ \frac{\rho}{2}(f_X(X)-f_Z(Z))^2\Bigr].
\label{eq:zero0}
\end{equation}

The first term is the usual prediction error loss, while
the second term is an ``agreement'' penalty,
encouraging alignment of predictions from different modalities.

To be more concrete in the setting of regularized linear regression, 
for a fixed
value of the hyperparameter $\rho\geq 0$, cooperative learning finds $\tx \in \mR^{p_x}$
and $\vec{\tz} \in \mR^{p_z}$ that minimize: 
\begin{multline}
J(\tx,\tz) =  
\frac{1}{2}  ||\y-X\tx- Z\tz||^2+ \frac{\rho}{2}||(X\tx- Z\tz)||^2 \\+
\lambda_x ||\tx||_1+ \lambda_z ||\tz||_1,
\label{eq:obj2}
\end{multline}
where $\rho$ is the hyperparameter that controls the relative importance of the agreement penalty term $||(X\tx- Z\tz)||^2$ in the objective,
and $\lambda_x ||\tx||_1$ and $\lambda_z ||\tz||_1$ are $\ell_1$ penalties\footnote{We assume that the columns of $X$ and $Z$ have been standardized, and $\y$ has mean 0 (hence we can omit the intercept). We use the commonly-used $\ell_1$ penalties for illustration, while the framework generalizes to other penalty functions.}. 

When $\rho=0$, cooperative learning reduces to early fusion, where we simply use the combined set of features in a supervised learning procedure.
When $\rho=1$, we can show that it yields a simple form of late fusion.
In addition, theoretical analysis has demonstrated that the agreement penalty offers an advantage in reducing the mean-squared error of the predictions under a latent factor model \citep{ding2022cooperative}.

\subsection{Semi-supervised cooperative learning}

In this section, we present \emph{semi-supervised cooperative learning}, which enables us to harness the power of both labeled and unlabeled data for multiomics data fusion.
Consider feature matrices $X \in \mR^{n \times p_x}$, $Z \in \mR^{n \times p_z}$, with labels $\y \in \mR^{n}$, and then additional feature matrices $X' \in \mR^{n_{\text{unlabeled}} \times p_x}$, $Z' \in \mR^{n_{\text{unlabeled}} \times p_z}$, without labels. 
The objective of semi-supervised cooperative learning is
\begin{align}
{\rm min} \;  {\rm E}\Bigl[\frac{1}{2}  (\y-f_X(X)-f_Z(Z))^2  
+\frac{\rho}{2}[(f_X(X)-f_Z(Z))^2 + (f_X(X')-f_Z(Z'))^2]\Bigr].
\end{align}


The agreement penalty allows us to use the matched unlabeled samples to our advantage, by encouraging predictions from different data modalities to align on both labeled and unlabeled samples, thus leveraging the aligned signals across omics in a semi-supervised manner.
This agreement penalty term is also related to ``contrastive learning'' \citep{chen2020simple, khosla2020supervised}, which is an unsupervised learning technique first proposed for learning visual representations.
Without the supervision of $\y$, it learns representations of images by maximizing agreement between differently augmented ``views'' of the same data example.
While contrastive learning is unsupervised and cooperative learning is supervised, both of which have a term in the objective that encourages agreement between correlated views, semi-supervised cooperative learning combines the strengths of both paradigms to fully exploit labeled and unlabeled data simultaneously.


In the regularized regression setting and with a common $\lambda$\footnote{It was shown in \cite{ding2022cooperative} that there is generally no advantage to allowing different $\lambda$ values for different modalities.}, the objective becomes
\begin{multline} 
J(\tx,\tz) =  
\frac{1}{2}  ||\y-X\tx- Z\tz||^2
+ \frac{\rho}{2}[||(X\tx- Z\tz)||^2 + ||(X'\tx- Z'\tz)||^2] \\+
\lambda( ||\tx||_1+  ||\tz||_1), 
\label{eq:sol_full2}
\end{multline}
and one can compute a regularization path of solutions indexed
by $\lambda$. 
Problem (\ref{eq:sol_full2})  is convex, and the solution can be computed  as follows. Letting \begin{equation}
\tilde X=
\begin{pmatrix}
 X   & Z\\
-\sqrt{\rho}X &  \sqrt{\rho}Z \\
-\sqrt{\rho}X' &  \sqrt{\rho}Z'
\end{pmatrix},  
\tilde \y=
\begin{pmatrix}
\y \\
\vec{0} \\
\vec{0}
\end{pmatrix}, 
\tilde \bbeta=
\begin{pmatrix}
\tx \\ \tz  
\end{pmatrix},
\label{eq:sol2}
\end{equation}
then the equivalent problem to \eqref{eq:sol_full2} is 
\begin{equation} 
\frac12||\tilde \y-\tilde X\tilde \bbeta||^2+\lambda( ||\tx||_1 +||\tz||_1) .
\label{eq:sol_full}
\end{equation}
This is a form of the lasso, and can be computed, for example
by the {\tt glmnet} package \citep{FHT2010}. 

Let ${\rm Lasso}(X,{\vec{y}},\lambda)$ 
 denote the generic problem:
\be
{\rm min}_{\bbeta} \;  \frac12\|{\vec{y}}- X\bbeta\|^2 +
\lambda \|\bbeta\|_1.
\ee
We outline the algorithm for semi-supervised cooperative learning in Algorithm \ref{alg:full_direct}.

\begin{algorithm}[h]
\caption{\em Semi-supervised cooperative learning.}
\label{alg:direct_alg}
\KwIn{$X \in \mR^{n\times p_x}$ and $Z \in
\mR^{n\times p_z}$, the response $\vec{y} \in \mR^{n}$, and the unlabeled data $X' \in \mR^{n_{\text{unlabeled}} \times p_x}$ and $Z' \in
\mR^{n_{\text{unlabeled}} \times p_z}$,
and a grid of hyperparameter values ($\rho_{\tt min}, \ldots, \rho_{\tt max}).$}

\For{$\rho \gets \rho_{\tt min},\hdots, \rho_{\tt max}$}{
     Set \begin{align*} \tilde X=
\begin{pmatrix}
 X   & Z\\
-\sqrt{\rho}X &  \sqrt{\rho}Z \\
-\sqrt{\rho}X' &  \sqrt{\rho}Z'
\end{pmatrix}, 
\tilde \y=
\begin{pmatrix}
\y \\
\vec{0} \\
\vec{0}
\end{pmatrix}.
\end{align*}
Solve ${\rm Lasso}(\tilde X, \tilde \y, \lambda$) over a decreasing grid of $\lambda$ values. 
}
\vspace{2mm}

Select the optimal value of $\rho^{*}$ based on the CV error and get the final fit.
\label{alg:full_direct}
\end{algorithm}

\section{Experiments}
\subsection{Simulated studies}

We first compare semi-supervised cooperative learning with vanilla cooperative learning, early and late fusion methods in simulation studies.
We generated Gaussian data with $n=200$ and $p=500$ in each of two views $X$ and $Z$, and created correlation between them using latent factors.
The response $\y$ was generated as a linear combination of the latent factors, corrupted by Gaussian noise.
We then generated an additional set of unlabeled data $X'$ and $Z'$ with $n_{\text{unlabeled}}=200$ and $p=500$.

\begin{figure*}[h!]  
\includegraphics[width=1.00\textwidth, clip]{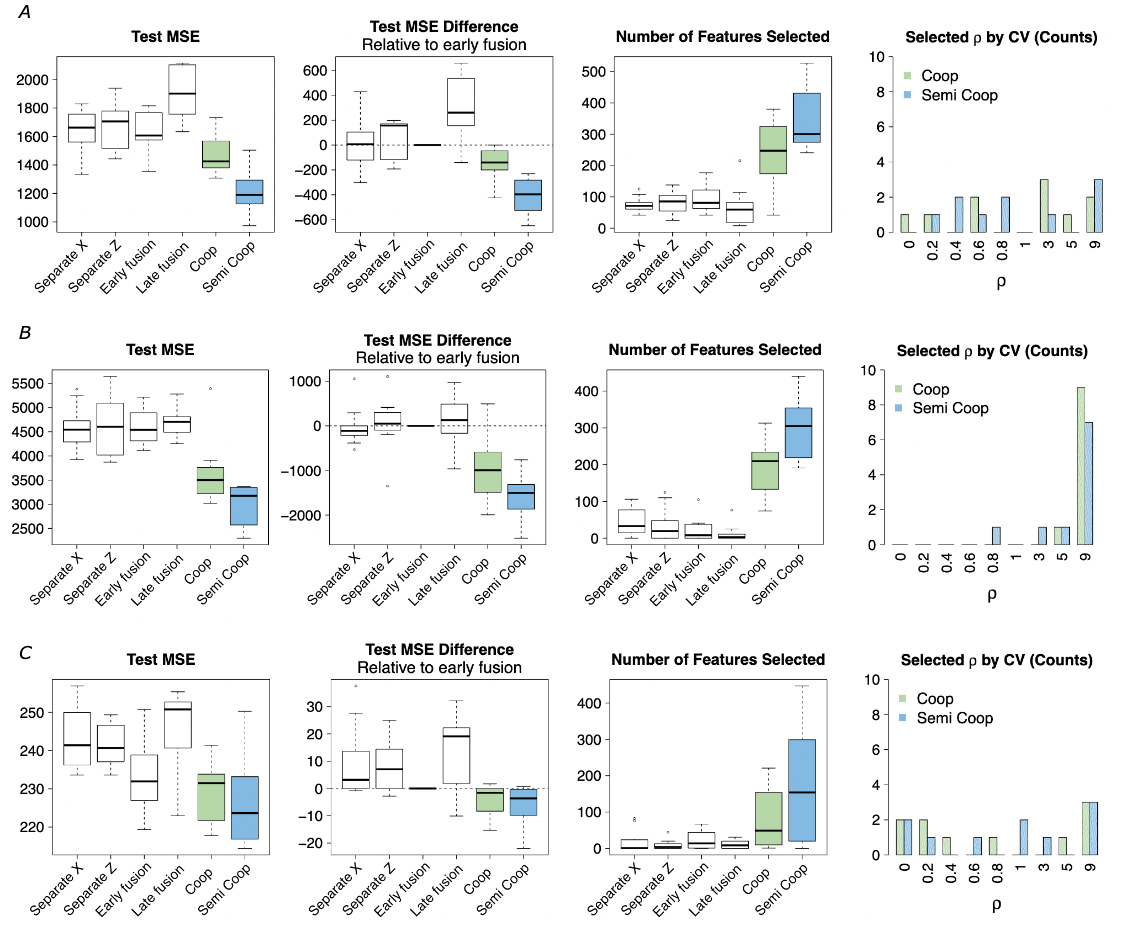}
\caption{\em Simulation studies on semi-supervised cooperative learning. {\emph(A)} Simulation results when $X$ and $Z$ have a medium level of correlation ($t = 2, s_{u} = 1$); both $X$ and $Z$ contain signal ($b_{x} = b_{z} = 2$), $n = 200, n_{\text{unlabel}} = 200, p = 1000$, SNR $= 1.8$.
The first panel shows MSE on a test set; the second panel shows the MSE difference on the test set relative to early fusion; the third panel shows the number of features selected; the fourth panel shows the $\rho$ values selected by CV in cooperative 
learning and semi-supervised cooperative learning. 
Here ``Coop'' refers to cooperative 
learning and ``Semi Coop'' refers to semi-supervised cooperative learning.
{\emph(B)} Simulation results when $X$ and $Z$ have a high level of correlation ($t = 6, s_{u} = 1$); only $X$ contains signal ($b_{x} = 2, b_{z} = 0$), $n = 200, n_{\text{unlabel}} = 200, p = 1000$, SNR $= 0.5$.
{\emph(C)} \em Simulation results when $X$ and $Z$ have no correlation ($t = 0, s_{u} = 1$); both $X$ and $Z$ contain signal ($b_{x} = b_{z} = 2$), $n = 200, n_{\text{unlabel}} = 200, p = 1000$, SNR $= 1.0$.}
\vspace{-3mm}
\end{figure*}

The simulation is set up as follows.
Given values for parameters $n, n_{\text{unlabled}}, p_{x}$, $p_{z}, p_{u}, s_{u}, t_{x}, t_{z}, \bbeta_{u}, \sigma$, we generate data according to the following procedure: 
\begin{enumerate}[noitemsep]
    \item $x_j \in \mR^{n}$ and $x'_j \in \mR^{n}$ distributed i.i.d. MVN$(0, I_{n})$ for $j = 1,2,\ldots,p_{x}$.
    \item $z_j \in \mR^{n}$ and $z'_j \in \mR^{n}$ distributed i.i.d. MVN$(0, I_{n})$ for $j = 1,2,\ldots,p_{z}$.
    \item For $i = 1, 2, \ldots, p_{u}$ ($p_{u}$ corresponds to the number of latent factors, $p_{u} < p_{x}$ and $p_{u} < p_{z}$):
    \begin{enumerate}[noitemsep]
        \item $u_{i} \in \mR^{n}$ and $u'_{i} \in \mR^{n}$ distributed i.i.d. MVN$(0, s_{u}^2I_{n})$;
        \item $x_{i} = x_{i} + t_{x} * u_{i}$, \quad $x'_{i} = x'_{i} + t_{x} * u'_{i}$;
        \item $z_{i} = z_{i} + t_{z} * u_{i}$, \quad $z'_{i} = z'_{i} + t_{z} * u'_{i}$.
    \end{enumerate}
    \item $X=[x_{1}, x_{2}, \ldots, x_{p_{x}}]$, $Z = [z_{1}, z_{2}, \ldots, z_{p_{z}}]$.
    \item $X'=[x'_{1}, x'_{2}, \ldots, x'_{p_{x}}]$, $Z' = [z'_{1}, z'_{2}, \ldots, z'_{p_{z}}]$.
    \item $U = [u_{1}, u_{2}, \ldots, u_{p_{u}}]$, $\y = U\bbeta_{u} + \epsilon$ where $\epsilon \in \mR^{n}$ distributed i.i.d. MVN$(0, \sigma^2 I_{n})$.
\end{enumerate}

There is sparsity in the solution since a subset of columns of $X$ and $Z$ are independent of the latent factors used to generate $\y$.
We use 10-fold CV to select the optimal values of hyperparameters.
We compare the following methods: (1) separate $X$ and separate $Z$ on the labeled data: the standard lasso is applied on the separate data modalities of $X$ and $Z$ with 10-fold CV; (2) early fusion on the labeled data: the standard lasso is applied on the concatenated data modalities of $X$ and $Z$ with 10-fold CV (note that this is equivalent to cooperative learning with $\rho = 0$); (3) late fusion on the labeled data: separate lasso models are first fitted on $X$ and $Z$ independently with 10-fold CV, and the two resulting predictors are then combined through linear least squares for the final prediction; (4) cooperative learning on the labeled data; (5) semi-supervised cooperative learning on both the labeled and unlabeled data\footnote{Traditional supervised learning models are not directly applicable to scenarios involving both labeled and unlabeled data.}.

Overall, the simulation results can be summarized as follows:
\begin{itemize}
    \item Semi-supervised cooperative learning performs the best in terms of test MSE across the range of SNR and correlation settings. It is most helpful when the data views are correlated and both contain signals, as shown in Figure 2{\em A}. 
    \item When there is no correlation between data views but each data view carries signals, semi-supervised cooperative learning still offers performance advantages as it utilizes the signals in both labeled and unlabled data, as shown in Figure 2{\em C}.
    \item When the correlation between data views is higher, higher values of $\rho$ are more likely to be selected, as shown in Figure 2{\em B} compared to Figure 2{\em A}. In addition, cooperative learning-based methods tend to select more features.
\end{itemize}

\subsection{Real data example}

We applied semi-supervised  cooperative learning to a real multiomics dataset of aging, collected from a cohort of 100 healthy individuals and individuals with prediabetes, as described in \cite{zhou2019longitudinal}.
Proteomics and transcriptomics were measured on the cohort: the proteomics data contained measurements for 302 proteins and the transcriptomics data contained measurements for 8,556 genes. 
The goal of the analysis is to predict age using proteomics and transcriptomics data and uncover molecular signatures associated with the aging process.

We split the data set of 100 individuals into training and test sets of 75 and 25 individuals, respectively.
We artificially masked the labels for half of the training samples to create a mix of labeled and unlabeled data.
Both the proteomics and transcriptomics measurements were screened by their variance across the subjects.
We averaged the expression levels for each individual across time points in the longitudinal study and predicted the corresponding age. 
We conducted the same set of experiments across 10 different random splits of the training and test sets.

\begin{table*}[t!]
    \footnotesize
    \caption{\emph{Multiomics studies on aging.} The first two columns in the table show the mean and standard deviation (SD) of mean absolute error (MAE) on the test set across different splits of the training and test sets; the third and fourth columns show the MAE difference relative to early fusion. The methods include (1) separate proteomics: the standard lasso is applied on the proteomics data only; (2) separate transcriptomics: the standard lasso is applied on the transcriptomics data only; (3) early fusion: the standard lasso is applied on the concatenated data of proteomics and transcriptomics data; (4) late fusion: separate lasso models are first fit on proteomics and transcriptomics independently and the predictors are then combined through linear least squares; (5) cooperative learning; (6) semi-supervised cooperative learning.}
    \begin{center}
    \begin{tabular}{c|M{1.45cm}M{1.45cm}|M{1.45cm}M{1.45cm}}
        \toprule
        \textbf{Methods} & \multicolumn{2}{c|}{\textbf{Test MAE}} & \multicolumn{2}{c}{\textbf{Relative to}} \\
        &  &  & \multicolumn{2}{c}{\textbf{Late Fusion}} \\
        & Mean & Std & Mean & Std  \\ \hline
        \midrule
        Separate Proteomics & 8.49 & 0.40 & -0.04 & 0.12 \\ 
        Separate Transcriptomics & 8.44 & 0.35 & -0.08 & 0.20 \\
        Early fusion & 8.52 & 0.32 & 0 & 0  \\
        Late fusion & 8.53 & 0.29 & 0.01 & 0.13\\
        \textbf{Cooperative learning} & 8.16 & 0.40  &  -0.37 & 0.16 \\
        \textbf{Semi-supervised cooperative learning} & \textbf{7.85} & \textbf{0.47} & \textbf{-0.67} & \textbf{0.26} \\
        \bottomrule
    \end{tabular}
    \end{center}
    \label{results}
\end{table*}

The results are shown in Table \ref{results}.
The model fit on the transcriptomics data achieves lower test MAE than the one fit on the proteomics data.
Early and late fusion hurt performance as compared to the model fit on only proteomics or transcriptomics.
Cooperative learning outperforms both early and late fusion by encouraging the predictions to align with each other.
Semi-supervised cooperative learning gives further performance gains by utilizing both the labeled and unlabeled data.
Moreover, it selects important features not identified by the other methods for predicting age, including PDK1, MYSM1, ATP5A1, APOA4, MST, A2M, which have been previously demonstrated to be associated with the aging process 
\citep{an2020inhibition, tian2020mysm1, choi2019c9orf72,goldberg2018mitochondrial, blacker1998alpha, garasto2003study, lee2013mst1, shang2022role}.

\section{Conclusion}
We introduce semi-supervised cooperative learning for multiomics data fusion in the presence of both labeled and unlabeled data.
By exploiting the shared underlying relationships across omics through an agreement penalty in both labeled and unlabeled data, our proposed approach demonstrates improved predictive accuracy on simulated studies and a real multiomics study of aging.
The agreement penalty allows us to effectively incorporate the unlabeled samples in the learning process and leverage them to our advantage. 
To our knowledge, our work represents a pioneering effort in multi-omics data fusion that unlocks the untapped potential of unlabeled data, enabling us to harness the valuable information that would otherwise remain unused for the discovery of novel insights and enhanced predictive modeling of diagnostics and therapeutics.




%
%
%
\bibliographystyle{unsrtnat}
\bibliography{mybibliography}

\end{document}